\documentclass[aps,prd,reprint,superscriptaddress,nofootinbib,longbibliography]{revtex4-1}

\usepackage{amssymb,amsmath,bm,natbib}
\usepackage[dvipsnames]{xcolor}
\usepackage[colorlinks = true,
            linkcolor = Maroon,
            urlcolor  = Maroon,
            citecolor = Maroon]{hyperref}
\usepackage{microtype}
\usepackage{graphicx}
\usepackage{xspace}
\usepackage{bbm}
\usepackage{tikz}
\usetikzlibrary{decorations.markings}
\usetikzlibrary{decorations.pathmorphing}
\tikzset{snake it/.style={decorate, decoration=snake}}
\usetikzlibrary{matrix,calc,arrows.meta}
\usepackage{float}
\usepackage{physics}

\renewcommand{\textit}[1]{#1}

\newcommand{\paral}{\ensuremath{{|\mkern-1mu|}}}

\makeatletter
\newenvironment{itemeq}
{%
  \par\noindent
  \hspace*{-0.1\leftmargin}%
  \begin{minipage}{\dimexpr\linewidth-\leftmargin+0.2em\relax}%
    \setlength{\displayindent}{0pt}%
    \setlength{\displaywidth}{\linewidth}%
    \@mathmargin=0pt\relax
    \tagshift@=\z@\relax
}
{%
  \end{minipage}\par
}
\makeatother

\usepackage{listings}
\usepackage{xcolor}

\definecolor{vsbg}{RGB}{30,30,30}        
\definecolor{vsbase}{RGB}{212,212,212}   
\definecolor{vskeyword}{RGB}{86,156,214} 
\definecolor{vscontrol}{RGB}{197,134,192}
\definecolor{vsbuiltin}{RGB}{220,220,170}
\definecolor{vstype}{RGB}{78,201,176}    
\definecolor{vscomment}{RGB}{106,153,85} 
\definecolor{vsstring}{RGB}{206,145,120} 
\definecolor{vsnumber}{RGB}{181,206,168} 
\definecolor{vslineno}{RGB}{133,133,133} 

\makeatletter
\newcommand{\jldigit}[1]{%
  \ifnum\lst@mode=\lst@Pmode\relax{\color{vsnumber}#1}\else#1\fi}
\makeatother

\lstdefinelanguage{Julia}{
  sensitive=true,
  alsoletter={!,?,@,_},
  keywords={function, global, local, const, struct, mutable, abstract,
      primitive, macro, module, let, quote, where, true, false, nothing},
  morekeywords=[2]{if, else, elseif, for, while, break, continue, return,
      do, begin, try, catch, finally, end, using, import, export, in, isa},
  morekeywords=[3]{rand, randn, sqrt, exp, log, sin, cos, sinh, acos, abs,
      floor, inv, mod, min, max, minimum, maximum, push!, sort!, mean,
      length, isempty, isfinite, enumerate, collect, union, get, range,
      quadgk, vegas, optimize, besseli, error, println, print, display,
      plot, plot!, scatter!, open, run, flush, redirect_stdout, abspath},
  morekeywords=[4]{@show, @time, @info, @threads, @__FILE__},
  morekeywords=[5]{Int, Float64, BigFloat, NTuple, Tuple, Vector, Array,
      Dict, Symbol, String, Bool},
  morecomment=[l]{\#},
  morecomment=[n]{\#=}{=\#},
  morestring=[s]{"""}{"""},
  morestring=[b]{"},
}

\lstdefinestyle{vscode-dark}{
  language          = Julia,
  basicstyle        = \small\ttfamily\color{vsbase},
  backgroundcolor   = \color{vsbg},
  keywordstyle      = \color{vskeyword},
  keywordstyle      = [2]\color{vscontrol},
  keywordstyle      = [3]\color{vsbuiltin},
  keywordstyle      = [4]\color{vsbuiltin},
  keywordstyle      = [5]\color{vstype},
  identifierstyle   = \color{vsbase},
  commentstyle      = \itshape\color{vscomment},
  stringstyle       = \color{vsstring},
  numbers           = left,
  numberstyle       = \tiny\color{vslineno},
  numbersep         = 8pt,
  frame             = none,
  framesep          = 4pt,
  xleftmargin       = 18pt,
  breaklines        = true,
  breakatwhitespace = true,
  showstringspaces  = false,
  tabsize           = 4,
  captionpos        = b,
  columns           = fullflexible,
  keepspaces        = true,
  literate=
    {0}{{{\jldigit{0}}}}1 {1}{{{\jldigit{1}}}}1 {2}{{{\jldigit{2}}}}1
    {3}{{{\jldigit{3}}}}1 {4}{{{\jldigit{4}}}}1 {5}{{{\jldigit{5}}}}1
    {6}{{{\jldigit{6}}}}1 {7}{{{\jldigit{7}}}}1 {8}{{{\jldigit{8}}}}1
    {9}{{{\jldigit{9}}}}1
    {λ}{{$\lambda$}}1 {ϵ}{{$\epsilon$}}1 {ε}{{$\varepsilon$}}1
    {θ}{{$\theta$}}1 {φ}{{$\varphi$}}1 {ϕ}{{$\phi$}}1 {α}{{$\alpha$}}1
    {π}{{$\pi$}}1 {μ}{{$\mu$}}1 {ρ}{{$\rho$}}1 {σ}{{$\sigma$}}1
    {χ}{{$\chi$}}1 {Γ}{{$\Gamma$}}1 {Δ}{{$\Delta$}}1 {Ω}{{$\Omega$}}1
    {∈}{{$\in$}}1 {≤}{{$\leq$}}1 {≥}{{$\geq$}}1 {≠}{{$\neq$}}1
    {≈}{{$\approx$}}1 {∝}{{$\propto$}}1 {→}{{$\rightarrow$}}1
    {⋅}{{$\cdot$}}1 {×}{{$\times$}}1 {√}{{$\sqrt{\,}$}}1
    {à}{{\`a}}1 {è}{{\`e}}1 {é}{{\'e}}1 {ì}{{\`i}}1 {ò}{{\`o}}1
    {ù}{{\`u}}1 {È}{{\`E}}1 {’}{{'}}1 {‘}{{'}}1 {“}{{``}}1 {”}{{''}}1
    {–}{{--}}1 {—}{{---}}1,
}
\begin{document}
\tighten  

\title{Hydrogenated carbon structures as directional sub-GeV dark matter detectors}

\author{Tom\'as~Arias}
\affiliation{Department of Physics, Cornell University, Ithaca, New York 14853, USA}

\author{Antonino~Bellinvia}
\affiliation{Dipartimento di Fisica, Sapienza Universit\`a di Roma, Piazzale Aldo Moro 2, I-00185 Rome, Italy}

\author{Gianluca~Cavoto}
\email{gianluca.cavoto@roma1.infn.it}
\affiliation{Dipartimento di Fisica, Sapienza Universit\`a di Roma, Piazzale Aldo Moro 2, I-00185 Rome, Italy}
\affiliation{INFN Sezione di Roma, Piazzale Aldo Moro 2, I-00185 Rome, Italy}

\author{Angelo~Esposito}
\email{angelo.esposito@uniroma1.it}
\affiliation{Dipartimento di Fisica, Sapienza Universit\`a di Roma, Piazzale Aldo Moro 2, I-00185 Rome, Italy}
\affiliation{INFN Sezione di Roma, Piazzale Aldo Moro 2, I-00185 Rome, Italy}

\author{Francesco~Pandolfi}
\email{francesco.pandolfi@roma1.infn.it}
\affiliation{INFN Sezione di Roma, Piazzale Aldo Moro 2, I-00185 Rome, Italy}

\author{Guglielmo~Papiri}
\email{gp343@cornell.edu}
\affiliation{Department of Physics, Cornell University, Ithaca, New York 14853, USA}

\author{Antonio~D.~Polosa}
\email{antonio.polosa@roma1.infn.it}
\affiliation{Dipartimento di Fisica, Sapienza Universit\`a di Roma, Piazzale Aldo Moro 2, I-00185 Rome, Italy}
\affiliation{INFN Sezione di Roma, Piazzale Aldo Moro 2, I-00185 Rome, Italy}

\author{Tyler~Wu}
\affiliation{Department of Physics, Cornell University, Ithaca, New York 14853, USA}

\date{\today}

\begin{abstract}
\noindent We propose hydrogenated carbon structures as targets with a remarkable sensitivity to dark matter--nucleon interactions, in the mass range between 1 MeV and 100 MeV. The ejection of a proton following the interaction with a dark matter particle is a quasielastic process, with an extremely small energy threshold, and a clear experimental signature. The proposed detectors are simple, technologically ready, and inexpensive. Yet, they can be considerably more sensitive than current experiments. They also allow strong directionality to be used toward efficient background rejection.
\end{abstract}

\maketitle


\section{Introduction}

\noindent Despite being one of the most compelling problems in fundamental physics, the search for  Weakly Interacting Massive dark matter Particles (WIMPs)  has so far only reported null results~\cite[e.g.,][]{PICO:2017tgi,SuperCDMS:2020aus,PandaX-4T:2021bab,DarkSide-50:2022qzh,SENSEI:2023zdf,LZ:2024zvo,CRESST:2024cpr,XENON:2025vwd}. This has fueled several efforts toward finding new ways to probe yet unexplored regions both in mass and couplings, as suggested by a number of theoretical models~\cite[e.g.,][]{Boehm:2003hm,Hooper:2008im,Feng:2008ya,Falkowski:2011xh,Hochberg:2014dra,DAgnolo:2015ujb,Kuflik:2015isi,DAgnolo:2018wcn,Essig:2011nj}. In particular sub-GeV dark matter particles cannot be detected in elastic scatterings with most of the detector targets, because of the large mass mismatch with the recoiling nuclei. Those experiments that are currently looking into the MeV to GeV range, or propose to do so, circumvent this limitation by trying to detect inelastic processes~\cite[e.g.,][]{Essig:2011nj,Kouvaris:2016afs,Chao:2021liw}, such as phonon emission in various materials~\cite[e.g.,][]{SuperCDMS:2020aus,SuperCDMS:2022kse,Colantoni:2020cet,vonKrosigk:2022vnf,CRESST:2024cpr,Griffin:2024jec,Angloher:2025fzw,TESSERACT:2025tfw,Bento:2025ijg} or the Migdal effect~\cite[e.g.,][]{XENON:2019zpr,DarkSide:2022dhx,SENSEI:2023zdf,PandaX-4T_2023} which, however, have much smaller rates, or by introducing light nuclei such as hydrogen in their targets~\cite[e.g.,][]{Bell:2023uvf,Amaro:2024vuk,NEWS-G:2024jms,HydroX:2025nxn}. 

We propose the use of {\it hydrogenated} carbon structures, such as graphene, graphite or nanotubes (CNTs), to search for dark matter particles interacting with nucleons. Indeed, when a sub-GeV dark matter particle impinges on the proton belonging to one of the hydrogen atoms, it can exchange enough energy to release it via an almost {\it elastic} process. Once emitted in a vacuum enclosure, the charged proton can be accelerated by an electric field, and eventually be collected. This way, the proposed concept does not rely on the detectability of a tiny nuclear recoil energy, but rather on the possibility of accelerating a charged particle up to detectable energies. Consequently, such a detector would have an extremely low energy threshold: as the typical binding energy of a proton onto a carbon structure is of the order of a few eVs~\cite[e.g.,][]{PTOLEMY:2022ldz,delfino2024multi,Casale:2025vgd}, dark matter particles with masses $m_\chi \sim \mathcal{O}({\rm MeV})$ can eject protons with sizable rates. In the case of vertically aligned CNTs, the expected signal is directional, thus providing a strong handle for background rejection. Graphene and CNTs have been studied both theoretically and experimentally to probe dark matter--electron interactions~\cite{Cavoto:2017otc,Hochberg:2016ntt,Pandolfi:2021tkx,Catena:2023awl,Catena:2023qkj,Das:2023cbv}, constrain the neutrino mass, and hunt the elusive relic neutrinos~\cite[e.g.,][]{PTOLEMY:2019hkd,PTOLEMY:2022ldz,Casale:2025vgd}. 

\begin{figure*}[t!]
    \centering
    \includegraphics[width=0.95\textwidth]{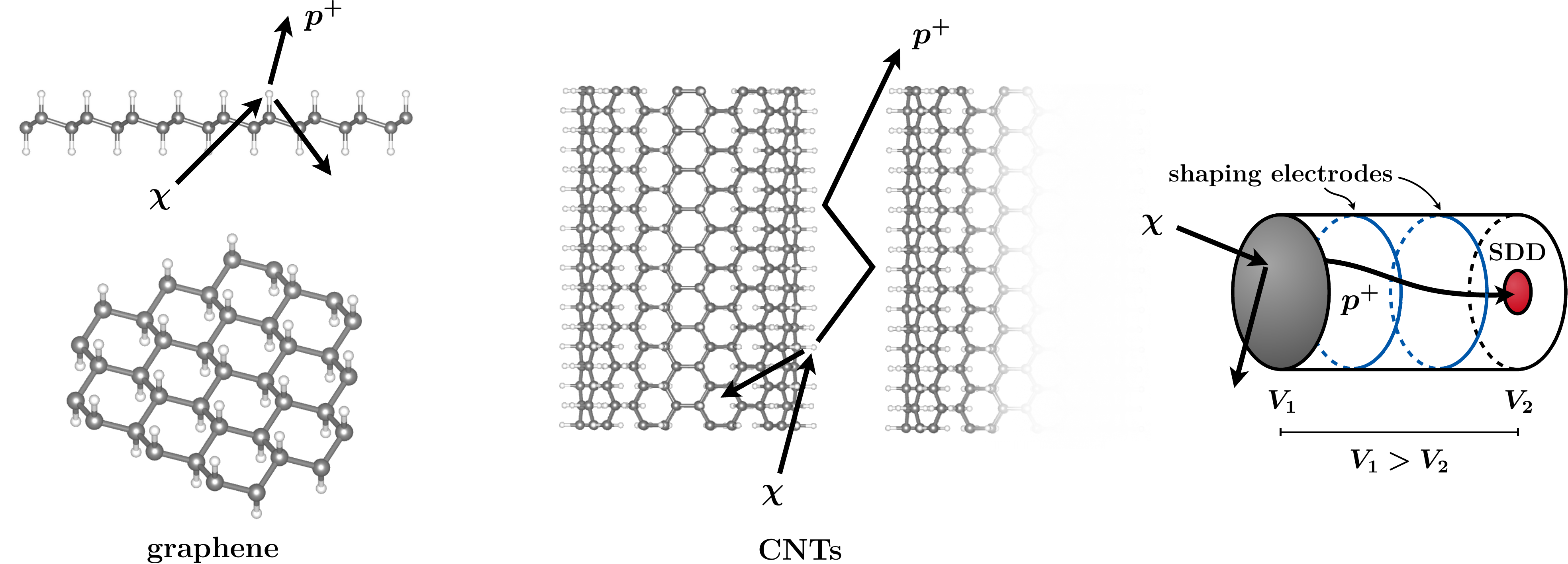}
    \caption{{\bf Left and central panels:} Schematic representation of the hydrogenated carbon structures at 100\% loading, and a pictorial dark matter interaction on a hydrogen nucleus, resulting in the ejection of a proton. Dark and light spheres represent carbon and hydrogen atoms, respectively, with the latter being covalently bound to their carbon site. {\bf Right panel:} Detector concept. After its ejection, the proton is accelerated by an electric field, and its trajectory is focused by shaping electrodes, until it is detected by an SDD.}
    \label{fig:scheme}
\end{figure*}

The technology behind the hydrogenation of carbon structures is  well assessed~\cite[e.g.,][]{abdelnabi2021deuterium,betti2022gap,tayyab2023spectromicroscopy,apponi2025highly,APPONI2026165658}, making it possible to engineer the supporting  graphene holder in order to optimize it toward the present proposal. Moreover, the proton binding energy can be varied by manipulating the hydrogenation level and/or the local geometry of the substrate~\cite[e.g.,][]{PTOLEMY:2022ldz,delfino2024multi,Casale:2025vgd}. Hydrogenated graphene is also stable at room temperature~\cite{Apponi:2025fmu}, making this detection scheme practically insensitive to thermal effects. This means that it can fit in a relatively small high-vacuum chamber, without the need for expensive infrastructures, such as cryostats. In the proposed detector, the ejected proton emerges from an anode, which is kept at a positive voltage of a few kVs. It can then be easily directed toward a sensor by a suitably shaped electric field.
This sensor, having a sensitive area much smaller than the hydrogenated substrate, can be a single-channel silicon drift detector (SDD). A SDD can also measure the accelerated proton energy with a resolution sufficient to separate signal from background~\cite{SIMSON2007772}. (See the Conclusion for a discussion on this.) Such a technique has already been implemented for the collection and detection of electrons, with large efficiencies of the order of $50\%$~\cite{Apponi:2022nxn}. Moreover, the components described above are all readily available technologies, further strengthening the case for the possibility of realizing this idea in the near future, and for a relatively contained cost.
The hydrogenated carbon structures considered here, as well as the proposed experimental concept, are shown in Figure~\ref{fig:scheme}. 

We consider a simple benchmark model of spin-independent dark matter--nucleon interactions, and employ Density Functional Theory (DFT)~\cite[e.g.,][]{kohn1965self,giustino2014materials} to determine the expected rate for the liberation of a proton from the substrate. We do this for both graphene and CNTs. The latter, in addition to offering substantially more target mass per unit area, are expected to exhibit directional sensitivity. In this case, however, one must also estimate the probability that the proton emitted by the dark matter scattering is transmitted to the exterior of the forest, without being lost or captured by the material itself. We do that with a suitably designed Monte Carlo simulation. In all these instances, we find that the proposed materials can outperform by orders of magnitude the best current  experimental constraints~\cite[e.g.,][]{SENSEI:2023zdf,PandaX-4T_2023}.


\section{The dark matter--proton collision}

\noindent We consider a benchmark model of fermionic dark matter interacting with protons via a spin-independent interaction mediated by a heavy scalar, $\mathcal{L}_{\rm int} = g_\chi \mkern2mu \phi \mkern2mu \bar \chi \chi + g_p \mkern2mu \phi \mkern2mu \bar p p$, where we assume $m_\phi \gg m_\chi v_\chi$, with $v_\chi$ being the typical dark matter velocity. From the quantum-mechanical viewpoint, this is matched to a contact interaction potential,\footnote{In the mass range of interest, the regime of contact interaction is valid for $m_\phi \gg \mathcal{O}(1\text{--}100) \text{ keV}$. For much larger masses, $m_\phi \gg \mathcal{O}(100) \text{ MeV}$, strong bounds are imposed by meson decays~\cite{Cox:2024rew,Cox:2025toz}, as well as by direct constraints on the mediators~\cite[e.g.,][]{Knapen:2017xzo}.}
\begin{align} \label{eq:Ucontact}
    U(\bm x_\chi, \bm x_p) = -\frac{g_\chi \mkern2mu g_p}{m_\phi^2} \, \delta(\bm x_\chi - \bm x_p) \,.
\end{align}
While we choose this model for concreteness and computational simplicity, we stress that a direct laboratory search is completely model independent, from the viewpoint of a specific dark matter model, as well as that of its cosmological history.

The initial state proton is bound to the carbon structure, and it is thus necessary to account for its wave function and binding energy. Realistic CNTs have a radius of order $10 \text{ nm}$, much larger than the size of a unit cell. Consequently, they can also be approximated as locally flat. We will assume a local coverage of 100\%, meaning that all carbon sites surrounding a given hydrogen atom are also occupied, and neighboring hydrogen atoms sit on opposite sides of the graphene layer.\footnote{Note that a {\it local} coverage of 100\% does not necessarily mean that all carbon sites of the {\it macroscopic} sample are hydrogenated.} This configuration, sometimes dubbed ``graphane,'' has been discussed in detail in~\cite{PTOLEMY:2022ldz,Casale:2025vgd} for the case of tritiated graphene. The initial proton wave function is
\begin{align} \label{eq:psi0}
    \psi_0(\bm x_p) = \frac{1}{\pi^{3/4} \lambda_\paral \sqrt{\lambda_{\perp} }} \, {\rm exp}\left(- \frac{x_{p,\paral}^2}{2\lambda_{\paral}^2} - \frac{x_{p,\perp}^2}{2\lambda_\perp^2} \right) \,,
\end{align}
where $\bm x_{p,\paral}$ and $x_{p,\perp}$ are the directions parallel and perpendicular to the graphene plane, respectively. Upon a rescaling from the tritium to the proton mass, the parallel and perpendicular spreads are $\lambda_{\paral} \simeq 0.17 \text{ \AA}$ and $\lambda_\perp \simeq 0.11 \text{ \AA}$, while the binding energy is $\varepsilon_0 \simeq{} - {} 4.5 \text{ eV}$~\cite{Casale:2025vgd}. 

In this work, we use graphene as a benchmark two-dimensional material, although large areas might be achieved more easily by leveraging the surface of graphite. The 100\% loading configuration described above, however, is not realizable on graphite, as the bottom parts of the carbon sites are bound to another graphene layer, and the hydrogenation is expected to happen all on the same side. A similar situation happens for CNTs. The 100\% loading configuration is only realizable on single-wall CNTs, which can be synthesized in lab~\cite[e.g.,][]{kumar2010chemical,chen2014chemical}. For the more common multiwall CNTs, hydrogenation happens only on the exterior of the tube. Nonetheless, the parameters characterizing the binding of the proton depend on the loading and spatial distribution only by, at most, a factor of order 1~\cite{PTOLEMY:2022ldz,delfino2024multi,Casale:2025vgd}. Our proposal and conclusions are thus unaffected.

For the processes of interest, the final state proton is free. Its wave function, as well as those of the incoming and outgoing dark matter, are then well approximated by plane waves. We normalize them to unity when working at some finite volume, $V$. The corresponding matrix element then reads,
\begin{align}
    \mathcal{M} \equiv \langle \bm k', \bm p | U | \bm k, \psi_0 \rangle = - \frac{g_\chi \mkern2mu g_p}{V^{3/2} \mkern1mu m_\phi^2} \, \hat \psi_0(\bm k - \bm k' - \bm p) \,,
\end{align}
where $\bm k$, $\bm k'$, and $\bm p$ are the momenta of the incoming dark matter, the outgoing one, and the outgoing proton, while $\hat \psi_0$ is the Fourier transform of the proton's initial wave function. Given this, the rate of ejection of a proton by dark matter is obtained from Fermi's golden rule:
\begin{align} \label{eq:dGamma}
    d\Gamma = 2\pi \, \big| \mathcal{M} \big|^2 \, \delta\left( \varepsilon_0 + E_k - E_{k'} - E_p \right) \frac{V d\bm k'}{(2\pi)^3} \frac{V d\bm p}{(2\pi)^3} \,,
\end{align}
where the $E$'s are kinetic energies, with obvious definitions. Finally, the expected experimental rate is obtained by accounting for the number of available protons $N_{\rm H}$, the mass density of incoming dark matter $\rho_\chi$, and its velocity distribution in the Milky Way halo $f(v)$. Specifically, it is given by
\begin{align} \label{eq:R}
    R(\hat{\bm v}_{\rm e}) = V N_{\rm H} \frac{\rho_\chi}{m_\chi} \int d\bm v \, f(|\bm v + \bm v_{\rm e}|) \, \Gamma(\bm v) \,,
\end{align}
where $\Gamma$ is the total rate obtained by integrating Eq.~\eqref{eq:dGamma}, which depends on the dark matter velocity through its initial momentum.
We take the velocity distribution to be a truncated Maxwellian as in the standard halo
model, with dispersion $v_0 = 230$~km/s, escape  velocity $v_{\rm esc} = 600$~km/s, and boosted with respect to the Galactic rest frame by the Earth velocity $v_{\rm e} = 240$~km/s~\cite{Piffl:2013mla,monari2018escape}. We also take the dark matter mass density to be $\rho_\chi = 0.4$~GeV/cm$^3$~\cite{Piffl:2013mla,ou2024dark}. We have also anticipated the possible directionality of our signal through the dependence of the rate on the direction of the Earth's velocity. The direction of the ``dark matter wind'' is opposite to $\hat{\bm v}_{\rm e}$.


\section{2D: graphene}

\noindent Eq.~\eqref{eq:R} provides the ideal rate for the liberation of a proton from the surface of the carbon structure. Yet, not all events are experimentally detectable. A possible obstacle lies in the fact that the proton could be ejected together with one of the electrons which were originally localized in its vicinity. This would produce a neutral hydrogen, which is much harder to detect.

We performed a DFT analysis to estimate the probability that the interaction ejects a naked proton, with no electrons around it. Given our flat graphene with a 100\% coverage, we computed the probability that, after the scattering, all electrons end up in the ground state of the new graphene sheet (identical to the initial one, but with one fewer proton). As described in detail in Appendix~\ref{app:initial_ejection}, we do this within the sudden approximation for the dark matter--proton interaction. This is valid for all masses of interest, except for those right around the MeV range, where corrections are expected. We conservatively find that the probability of ejecting a naked proton is
\begin{align} \label{eq:Pproton}
    P_{\rm proton} \gtrsim 72\% \,.
\end{align}

As far as graphene is concerned, after the proton is liberated without electrons bound to it, it is accelerated with a suitable electric field, and eventually collected by the detector---see again Figure~\ref{fig:scheme}.  The final rate is then
\begin{align}
    R_{\rm graphene}(\hat{\bm v}_{\rm e}) = R(\hat{\bm v}_{\rm e}) \, P_{\rm proton} \,.
\end{align}
We report the corresponding projections in Figure~\ref{fig:projections}, computed in a configuration with the dark matter wind orthogonal to the graphene layer. The strength of the dark matter--proton interaction is represented with a reference cross section, $\bar\sigma_p \equiv g_\chi^2 \mkern2mu g_p^2 \mkern2mu \mu_{\chi p}^2/(\pi \mkern1mu m_\phi^4)$, where $\mu_{\chi p}$ is the dark matter--proton reduced mass. We show two instances: a conservative one considering a graphene sheet of area $100 \text{ cm}^2$, and a more ambitious one with an area of $1 \text{ m}^2$. Assuming that our graphene sample is globally fully hydrogenated, these correspond respectively to a target mass of $M_{\rm H} \simeq 0.66$ and $66 \text{ }\mu\text{g}$. As one can see, already with this simple setup, the expected sensitivity is considerably better than the current experimental bounds. The minimal mass that can be probed by this detector is dictated by the initial proton binding energy, which must be overcome by the dark matter kinetic energy. This leads to $m_\chi^{\rm min} = 2 |\varepsilon_0|/(v_{\rm esc} + v_{\rm e})^2 \simeq 1.1 \text{ MeV}$.

\section{3D: nanotubes} \label{sec:CNTs}

\noindent Along similar lines, one can consider hydrogenated CNTs. This allows for a three-dimensional target, thus gaining considerably more target mass if compared with a graphene/graphite setup with equal surface area. Arrays of aligned CNTs have already been considered as a possible target for direct dark matter detection. 
In particular, extensive work has been devoted to trying to employ them to hunt for possible dark matter--electron interactions~\cite{Cavoto:2017otc,Hochberg:2016ntt,Pandolfi:2021tkx,Catena:2023awl,Catena:2023qkj}.
As far as the dark matter--nucleon coupling is concerned, they were originally considered to look for processes where the dark matter ejects a carbon nucleus~\cite{Capparelli:2014lua,Cavoto:2016lqo}. In that instance, however, the large carbon mass allowed one to probe masses no lighter than a few GeVs. 

In our case, instead, the target is a proton. Once emitted, the proton must travel to the outside of the array of CNTs to be eventually detected. Before doing that, however, the vast majority of the  protons scatter multiple times on the surfaces of the CNTs. It is therefore necessary to understand the result of these interactions: whether they scatter off elastically, inelastically, or if they get captured and lost.

A number of recent experimental results have investigated the behavior of protons impinging on the surface of graphene which, we recall, is locally very similar to single-wall CNTs. Among other things, these data show that (a) graphene is somewhat permeable to protons with kinetic energies above about 1~eV~\cite[e.g.,][]{hu2014proton,lozada2018giant,griffin2020proton,zeng2021biomimetic,wahab2023proton}, and (b) proton absorption on the layer is strongly suppressed when their kinetic energy is below about 0.2~eV~\cite{tong2024control}.\footnote{This behavior differs from that of other ions, which instead are blocked and absorbed by single layer graphene~\cite[e.g.,][]{hu2014proton,lozada2018giant,griffin2020proton,zeng2021biomimetic,wahab2023proton,tong2024control}. We also point out that early calculations suggested that protons are neutralized by the interaction with solid surfaces~\cite[e.g.,][]{horiguchi1978auger,Winter:Auger:1990,Jouin:Auger:2011,ZIMNY:Auger:1991}.  
Such results seem to be at odds with the above mentioned experiments. Nonetheless, we notice that they are obtained for proton kinetic energies larger than 100~eV, far from the regime discussed here.} Since the degree of permeability is hard to quantify at this stage, we adopt a conservative assumption: any proton with energy in the plane orthogonal to the CNT axis, $E_{xy}$, exceeding $E_{xy} > 0.2 \text{ eV}$ undergoes some inelastic processes, loses energy, and is eventually captured. Only protons with $E_{xy} < 0.2 \text{ eV}$ scatter elastically off the walls of the CNTs.
With this at hand, there are three possible fates for one of the protons ejected by the dark matter:
\begin{itemize}
    \item If its energy is $E_{xy} > 0.2 \text{ eV}$, it gets captured after interacting with the CNT wall.
    \item If its energy is $E_{xy} < 0.2 \text{ eV}$, but it leaves the array from either the bottom or from the sides, preventing its collection and observation.
    \item If its energy is $E_{xy} < 0.2 \text{ eV}$, and it manages to leave the array from the top.
\end{itemize}
Among these possibilities, only the last one is favorable to our setup. We indicate its probability as $P_{\rm exit}(\hat{\bm v}_{\rm e})$, anticipating the fact that, as shown below, this depends on the direction of the dark matter wind.

\begin{figure}
    \centering
    \includegraphics[width=\columnwidth]{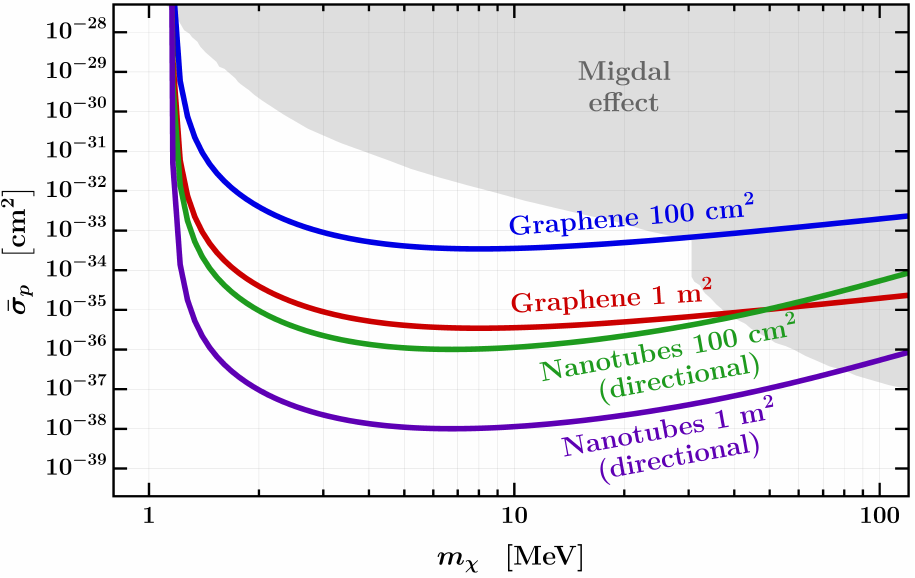}
    \caption{Projected reach at 90\%~C.L., assuming no background and 1 year of exposure. The target masses corresponding to graphene of area $100 \text{ cm}^2$ and $1 \text{ m}^2$ are $M_{\rm H} = 0.66$ and $66 \text{ }\mu\text{g}$, respectively. For CNTs of area $100 \text{ cm}^2$ and $1 \text{ m}^2$, they are instead $M_{\rm H} = 0.84$ and $84$~mg. The gray shaded areas correspond to currently excluded region, as obtained using the Migdal effect, whose leading bounds are set by SENSEI~\cite{SENSEI:2023zdf} and PandaX-4T~\cite{PandaX-4T_2023}. The reference cross section is defined as $\bar\sigma_p \equiv g_\chi^2 \mkern2mu g_p^2 \mkern2mu \mu_{\chi p}^2/(\pi \mkern1mu m_\phi^4)$.}
    \label{fig:projections}
\end{figure}

\begin{figure}[t]                       
    \centering
    \includegraphics[width=\columnwidth]{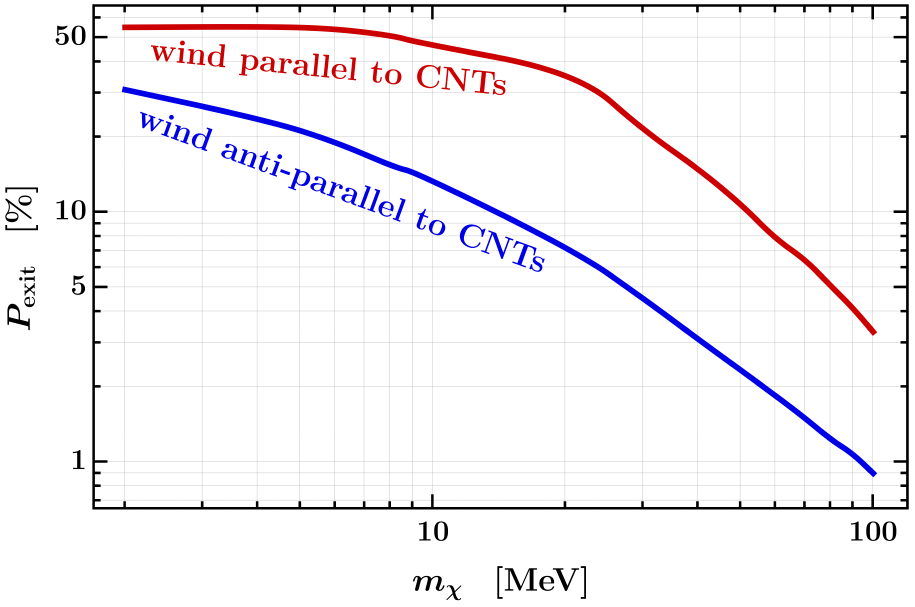}
    \par\vspace{0.5em}
    \caption{Probability for a proton to leave from the top side of the array of CNTs as a function of the dark matter mass, computed for a dark matter wind parallel ({\bf red curve}) and anti-parallel ({\bf blue curve}) to the orientation of the CNTs.}
    \label{fig:probs}
\end{figure}

To determine the fraction of protons that manage to leave the CNT forest from the top, we run a suitably designed Monte Carlo simulation, whose details are reported in Appendix~\ref{app:MC}. First of all, one of the main outcomes of our simulations is the following: for arrays of area larger than $0.5 \text{ mm} \times 0.5 \text{ mm}$, events lost by later exits are completely negligible. This result is central, as it remains valid even if one were to consider a more realistic CNT forest, where the spacing between CNTs is somewhat irregular. This conclusion is stronger for larger arrays, as it becomes much more likely for a proton to reach the top before reaching the sides. The results for the probability that a proton escapes from the top are reported in Figure~\ref{fig:probs}.
This probability decreases strongly with increasing dark matter masses. Indeed, higher masses imply a much larger chance for the outgoing proton to have $E_{xy} > 0.2 \text{ eV}$, and thus to fall among those events that we consider as lost.
From the large difference between the two curves, we also deduce that we expect order one modulations, when the direction of the dark matter wind with respect to the orientation of CNTs changes.
This can be used as powerful leverage to discriminate the signal from the (isotropic) background. The modulation decreases for masses $m_\chi \lesssim 10 \text{ MeV}$. This is indeed the regime where the typical dark matter momentum becomes smaller than the inverse spread of the initial proton wave function, $m_\chi v_\chi \lesssim 1/\lambda_{\paral},1/\lambda_{\perp}$---see Eq.~\eqref{eq:psi0}. Momentum conservation is then maximally broken by the wave function, and the final state proton retains less information about the direction of the incoming dark matter.

In light of all this, the expected event rate for an array of CNTs is given by
\begin{align}
    R_{\rm CNT}(\hat{\bm v}_{\rm e}) = R(\hat{\bm v}_{\rm e}) \, P_{\rm proton} \, P_{\rm exit}(\hat{\bm v}_{\rm e}) \,,
\end{align}
where $P_{\rm proton}$ is the probability reported in Eq.~\eqref{eq:Pproton}.
The corresponding projected sensitivity is again shown in Figure~\ref{fig:projections}. Specifically, we assume the CNTs to have a diameter of 10~nm, a distance of 50~nm between their centers, and a height of 100~$\mu$m~\cite{D5NR02221E}. For arrays with area $100 \text{ cm}^2$ and $1 \text{ m}^2$, the corresponding target masses are $M_{\rm H} = 0.84$ and $84$~mg, respectively.


\section{Conclusion}

\noindent We proposed the use of hydrogenated graphene, possibly graphite, and carbon nanotubes as targets to probe dark matter--nucleon interactions for masses as light as $m_\chi \sim\mathcal{O}(\rm MeV)$. Thanks to the weak carbon--hydrogen bond, and the quasielastic nature of the interactions, this setup leverages the net charge of the emitted proton, which allows us to accelerate it up to detectable energies, thus evading the limits posed by the experimental energy threshold on traditional nuclear recoil. For this reason, the proposed detector would be characterized by a small energy threshold, and a detection scheme based on these materials can be orders of magnitude more sensitive than current experiments.

The setup is expected to be inexpensive and easy to operate. Hydrogenated graphene and graphite are simple systems, with well-understood properties, thus allowing us to envision the realization of this proposal in a short timescale and with a reasonable opportunity to scale up the instrumented target mass. Moreover, CNTs provide substantially more target mass than an equivalent area of graphene/graphite. They are also expected to provide a strongly directional signal, offering a key tool toward background rejection. Nonetheless, they require better understanding, especially regarding the capture probability of protons scattering off their walls, the characterization of irregularities in their distribution and geometry~\cite{ripanti2019polarization}, and how these impact the proposed scheme.

Dark matter detectors usually require  good shielding against environmental radioactivity and cosmic rays, and they must be fabricated with  weakly radioactive materials. 
For instance, gamma rays can interact with the carbon atoms of the substrate and generate background Compton electrons reaching the SDD, and mimicking a dark matter signal. Compton electrons, in fact, can have energies as large as $100 \text{ keV}$, and are therefore not stopped by the applied electric field. Nonetheless, we calculated that, for the $100 \text{ cm}^2$ hydrogenated graphene target, it is sufficient to keep the gamma ray flux below a few Hz to have negligible background events in one year of exposure. A further background suppression technique might be based on the use of a localized magnetic field  to deflect electrons. Environmental neutrons are also expected to be innocuous. For instance, 
the typical neutron flux at the underground Gran Sasso laboratory is $\Phi \sim 10^{-6} \text{ neutrons/cm}^2\text{/s}$~\cite{Wulandari:2003cr}. The neutron--proton cross section is a function of energy. However, even considering only its highest possible value (achieved at very low energy), this is about $\sigma_{np} \sim 20 \text{ barns}$~\cite{Babenko:2007ss}. A simple calculation shows that the corresponding rate of background events due to environmental neutrons is always completely negligible, except for the largest CNT array considered here, of area $1 \text{ m}^2$. In that case, the background rate is expected to be $\mathcal{O}(10) \text{ events/year}$. Nonetheless, even in this instance, the neutron flux can easily be reduced by several orders of magnitude with proper shielding, as shown, for example, in~\cite{Stewart:2006cr}. Moreover, signal and background can be discriminated thanks to directionality.

This  work suggests a number of further theoretical and experimental investigations. A more accurate theoretical characterization of the materials will come, together with a more refined evaluation of the proton ejection probability, and with the study of different dark matter--proton couplings. These include light mediators and spin-dependent interactions, with the latter allowed thanks to the proton net spin. From the experimental viewpoint, the proposed mechanism of proton ejection can be validated by mimicking the dark matter particles with epithermal neutrons, with energies in the eV range.


\begin{acknowledgments}
\noindent We are grateful to Michele~Tarquini for collaboration at the early stages of this project. AE acknowledges support by the Italian Ministero dell’Universit\`a e della Ricerca through the FIS2 Starting Grant project LEAPS (CUP: B53C25003020001). TW was supported by the US National Science Foundation under award PHY-1549132, the Center for Bright Beams.
\end{acknowledgments}

\appendix
\section{Proton's initial ejection probability} \label{app:initial_ejection}

\noindent In principle, the final state of the initial dark matter scattering can be complicated: when a proton is ejected from the hydrogenated carbon structure, the electrons in the lattice can be promoted to excited lattice states, they can be ionized and freed from the lattice, or they can be ejected as bounded to the outgoing proton. In this appendix we use DFT to estimate the probability for the final state particle to be a naked proton, without electrons bound to it.

As far as the lattice structure is concerned, we consider a graphene layer with a hydrogen loading of $100\%$. This corresponds to a configuration where each hydrogen is bound to a carbon site, on alternating sides of the graphene sheet. Due to their large radius, this provides a good description of realistic CNTs as well. The CNTs we consider here, in fact, have a typical radius of the order of tens of nanometers, which is much larger than the size of the graphene unit cell.

We start by using DFT to compute the initial electronic state. We initialize our graphene supercell geometry and place hydrogen atoms regularly on the sheet until we reach the desired hydrogen coverage. Then, we allow the lattice to relax to its equilibrium configuration, so that the entire electrons--nuclei system is in its ground state. Using the single-particle wave functions from Kohn--Sham DFT~\cite[e.g.,][]{kohn1965self,giannozz_01}, we can construct the effective non-interacting electronic ground state as a Slater determinant for the supercell geometry, which we denote as $\Psi_{0}(\bm{x}_1, \dots ,\bm{x}_N)$, where $\bm{x}_i$ are the coordinates of the electrons. We consider this to be the electronic initial state.

To simplify the calculation, we only compute the probability for the event where, after the proton has been ejected, all the electrons remain bound to the lattice, in the original electronic ground state. This is only one of many possible final states. Indeed, as long as the proton is ejected naked, the electrons could end up in any possible excited state of the graphene system, without affecting our detection strategy. The probability computed here, therefore, provides a safe lower bound on the total detection rate achieved in a realistic experiment.

Following the so-called sudden approximation, and the approach developed for the determination of the Migdal effect rate \cite{Migdal1977Qualitative}, the probability that all electrons remain bound to the graphene sheet after a dark matter--proton scattering event is essentially given by the overlap between the initial ground state total electronic wave function (when the lattice is at equilibrium structure) and the final total electronic wave function when the lattice is modified by removing one proton from its structure.

To be more precise there are three different time scales in this process. The shortest time scale is the one associated to the dark matter--proton interaction. This can be estimated as the time it takes for the dark matter wave function to separate from that of the proton, 
\begin{align}
    \tau_\chi \sim \lambda_\chi/v_\chi = \frac{1}{m_\chi v_\chi^2} \lesssim 10^{-15} \text{ s} \,,
\end{align}
where $\lambda_\chi$ is the dark matter de Broglie wavelength, and we used the fact that we are interested in masses $m_\chi \gtrsim 1$~MeV. For masses larger than this, the typical scattering time, $\tau_\chi$, becomes smaller than $10^{-15} \text{ s}$. The time scale associated to the electronic response is, instead, roughly given by,
\begin{align}
    \tau_{e} \sim 1/\varepsilon_e \sim 10^{-15} \text{ s} \,,
\end{align}
where $\varepsilon_e$ is the electrons' characteristic energy. This can be taken to be of the order of the work function of the hydrogenated graphene system, that is the energy to remove one electron from the system, $\varepsilon_e \simeq 5 \text{ eV}$.
Thus, for all masses of interest, the scattering process happens faster than the electronic response, thus justifying the use of the sudden approximation---modulo possible corrections for masses close to the MeV.

In addition, the characteristic time of the electron response is much shorter than the time scale of the nuclear degrees of freedom of the lattice, $\tau_L$. This guarantees the applicability of the Born--Oppenheimer approximation which is at the basis of standard DFT methods. 

To compute the final electronic state, we remove one proton from the sheet without, however, removing its electron. The total number of electrons is thus unchanged. Given the separation of scales discussed above, we do not allow for the lattice to relax, as the electronic degrees of freedom respond much more quickly. We then solve for the new electronic ground state, the one obtained from a lattice with one fewer proton. The corresponding wave function is again obtained as a Slater determinant, which we denote as $\Psi'_{0}(\bm{x}_1,\dots,\bm{x}_N)$. 
The probability of ending up in the new ground state, as determined within the sudden approximation, is then simply,
\begin{align} \label{eq:ejection_prob}     P_{\text{ground}\mkern2mu\to\mkern2mu\text{ground}'}=\big|\bra{\Psi_0}\ket{\Psi_0'}\big|^2 \simeq 72\% \,.
\end{align}
As discussed before, this represents a conservative lower bound to the actual probability of emission of a naked proton.

For this calculation, we simulated a supercell geometry corresponding to a finite hydrogenated graphene flake, finding a fast numerical convergence in the calculation up to the size of $7\times7$ unit cells. Additional details on the DFT computation can be found in the next appendix.

\section{Computational details} \label{app:computational_details}

\noindent For the DFT calculations, we used JDFTx, an open-source DFT software, for the Kohn--Sham eigenstates and energies~\cite[]{SUNDARARAMAN2017278}. In the calculations, we use the Perdew--Burke--Ernzerhof generalized-gradient (gga-PBE) functional, along with the SG15 norm-conserving pseudopotentials~\cite[]{PhysRevLett.77.3865,SCHLIPF201536}. The plane wave cutoff we used was $30$ Hartrees. To limit periodic image interactions for planar and isolated geometries, we used Coulomb truncation~\cite[]{PhysRevB.87.165122}. 

For the calculation of the initial proton ejection probability, we used two hydrogenated graphene flakes consisting of $N$-by-$N$ primitive cells where the second flake has a hydrogen atom missing near the center. Letting $N=4,\,5,\,6,\,7$, we compute the probabilities and extrapolate the converged value as $N\rightarrow\infty$. To limit edge effects, we decorate the boundary of the flake with hydrogen atoms at appropriate carbon--hydrogen bonding sites. The length of the box in the normal direction is set to 30 Bohrs. Since the system is meant to be isolated in vacuum, we used Gamma-point-only Brillouin-zone sampling, i.e. we retained only Bloch states with zero crystal momentum; this one–k-point approximation is commonly used for large supercells, where the Brillouin zone becomes correspondingly small \cite{Martin2004ElectronicStructure}. The states are then indexed by a single integer. Using Slater determinant states for the electron ground states, the probability in Eq.~\eqref{eq:ejection_prob} is obtained from,
\begin{align}
    \langle \Psi_0 | \Psi_0' \rangle = {\rm det}\big( \langle \psi_m' | \psi_n \rangle \big) \,.
\end{align}
Here the determinant is computed of the $m$ and $n$ indices labeling the single-electron states. Moreover, $\psi$ and $\psi'$ are the states computed, respectively, before and after the removal of the proton.

\begin{table*}[t!]
    \centering
    \begin{tabular}{l|c|c|c|c|c|c|c|c}
        \hline\hline
        & \multicolumn{4}{c|}{\textbf{Wind parallel}} & \multicolumn{4}{c}{\textbf{Wind anti-parallel}} \\\hline\hline
        $m_\chi$ [MeV] & 5 & 10 & 50 & 100 & 5 & 10 & 50 & 100 \\\hline
        $E_{xy} > 0.2 \text{ eV}$ events \; & 0.23 & 0.39 & 0.87  & 0.95 & 0.23  & 0.38 & 0.87 & 0.95 \\
        Top exits &0.55 & 0.47 & 0.11 & 0.04 & 0.21 & 0.14  & 0.02 & 0.01 \\ 
        Bottom exits &0.21 & 0.14 & 0.02 & 0.01 & 0.56  & 0.48 & 0.11 & 0.04 \\
        Lateral exits &$0.01$& $<0.01$ &  $<$ 0.01 & $< 0.01$ & $<0.01$ & $<0.01$ & $<0.01$ & $<0.01$ \\
        \hline\hline
    \end{tabular}
    \caption{Fraction of events falling in the different categories described in the text, evaluated for a total of $N = 10^5$ protons, and for two relative directions of the dark matter wind with respect to the CNTs: parallel and anti-parallel.}
    \label{tab:MC}
\end{table*}

\section{Simulation of the fate of the proton} \label{app:MC}

\noindent The trajectory and scattering off the walls of the CNTs are simulated with Monte Carlo methods implemented in {\tt Julia} and {\tt Mathematica}. Specifically, {\tt Julia} models the detector geometry and proton transport in the CNT forest, while {\tt Mathematica} treats the dark matter--proton interaction and samples the proton’s initial velocity.

We model a detector consisting of a square grid of  $10^4$ CNTs per side, 100 $\mu $m in height, 10 nm in diameter, and with 50 nm spacing between the centers of adjacent CNTs, making the array area $0.5 \text{ mm} \times 0.5 \text{ mm}$. To speed up the simulation, the full grid is simulated by applying periodic boundary conditions to a single square cell with two CNTs per side.

The simulation consists in the following main blocks.

\begin{enumerate}
    \item Random sample of the dark matter velocity. This is done by an accept-reject method, from a distribution given by a truncated Maxwellian following the standard halo model, as described in the main text.

    \item Given the dark matter velocity, we extract the momentum of the ejected proton, as dictated by the differential distribution obtained from Eq.~\eqref{eq:dGamma}. Due to bad numerical convergence, for masses $m_\chi > 10 \text{ MeV}$, we approximate the proton wave function as isotropic, $\lambda_\paral = \lambda_\perp$. For large exchanged momenta, anisotropic effects are anyway negligible. The two distributions are the following:
    \begin{itemeq}
        \begin{subequations}
        \begin{align}
            \frac{d\Gamma_{\rm isot}}{d{\bm p}} \propto{}&
            \frac{e^{-\lambda^2(q^2+k'^2)}}{q}
            \sinh\!\left(2 \lambda^2 \mkern1mu q \mkern1mu k'\right) \,, \\
            \frac{d\Gamma_{\rm anisot}}{d\bm p} \propto{}&
            k' \int_{-1}^1 d\mkern-2mu\cos\theta \, I_0 \!\left( 2 \lambda_\paral^2 q_\paral k' \sin\theta \right) \notag \\
            {}& \times e^{-\lambda_\paral^2\left( q_\paral^2 + k^{\prime2}\sin^2\theta \right)} \\
            {}& \times e^{-\lambda_\perp^2 \left( q_\perp^2 + k^{\prime2}\cos^2\theta - 2 q_\perp k' \cos\theta \right)} \,, \notag
        \end{align}
        \end{subequations}
    \end{itemeq}
    where, for the isotropic distribution, we took $\lambda \equiv (\lambda_\paral +\lambda_{\perp}) / 2$, and we defined 
    $\bm q = \bm k - \bm p$, with $\bm k$ and $\bm p$ the momenta of the incoming dark matter and outgoing proton, respectively. Moreover, $q_{\perp}$ and $\bm q_{\paral}$ are the components orthogonal and parallel to the surface of the CNT, and $k'= \sqrt{2m_\chi \left( \varepsilon_0  + \frac{k^2}{2m_\chi} - \frac{p^2}{2m_p} \right)}$, with $m_p$ the proton mass. Finally, $I_0$ is a modified Bessel function of the first kind.
    
    \item If $E_{xy} > 0.2 \text{ eV}$ {\it or} the proton is emitted downward, the event is discarded. In the first instance it is assumed that it undergoes (yet unknown) energy losses and eventually gets captured. In the second instance, instead, it is guaranteed that it will never leave the forest from the top.

    \item If $E_{xy} < 0.2 \text{ eV}$ {\it and} the proton is emitted upward, we have two possible instances:
    \begin{itemize}
        \item If the proton is ejected toward the interior of the CNT, it can only undergo multiple scatterings inside the CNT and eventually leave from the top.

        \item If the proton is ejected outside the CNT, instead, it will scatter off the walls of the other CNTs in the forest. We then simulate its full trajectory. At each step, we determine which CNT it interacts with, compute the collision point and extract the vector normal to the surface of the CNT. After the collision, we flip the sign of the component of the proton's momentum parallel to this vector, while preserving the orthogonal components. Once the proton trajectory has been simulated there are two possible outcomes: the proton leaves the forest from one of the sides ({\it lateral exit}), or from the top ({\it top exit}). Of all these instances, only top exits are favorable under our detection scheme.
    \end{itemize} 
\end{enumerate}

We repeat this for $N = 10^5$ protons, and for two possible directions of the CNTs: parallel and anti-parallel to the dark matter wind. In Table~\ref{tab:MC} we report the results of our simulations for a few values of the dark matter mass. Notice that the probability that the initial proton is emitted with $E_{xy} < 0.2 \text{ eV}$ and directed upward is nothing but $P_{\rm top} + P_{\rm lateral}$, as these are the only two possible outcomes once the proton starts moving upward and scatters elastically. Given this, the probability for a proton to leave the array is,
\begin{align} \label{eq:Pexit}
    \begin{split}
        P_{\rm exit} ={}& \frac{1}{2} \left( P_{\rm top} + P_{\rm lateral} \right) + \frac{1}{2} P_{\rm top} \\
        \simeq{}& P_{\rm top} \,.
    \end{split}
\end{align}
The first term in the first line is the probability corresponding to the instance where the dark matter hits a proton on the inside of the CNT, while the second term is when it hits a proton on the outside. Finally, in the second line we used $P_{\rm lateral} \ll P_{\rm top}$, as shown in Table~\ref{tab:MC}. Note that, if realistic CNTs were to be capped at their top, this would result in a reduction of the number of observed events coming from protons initially bound to the interior of the CNTs. In the worst case scenario, this decreases the probability in Eq.~\eqref{eq:Pexit} by a factor of 1/2, which does not affect the quality of the proposal.

\bibliography{ref.bib}

@article{PandaX-4T:2021bab,
    author = "Meng, Yue and others",
    collaboration = "PandaX-4T",
    title = "{Dark Matter Search Results from the PandaX-4T Commissioning Run}",
    eprint = "2107.13438",
    archivePrefix = "arXiv",
    primaryClass = "hep-ex",
    doi = "10.1103/PhysRevLett.127.261802",
    journal = "Phys. Rev. Lett.",
    volume = "127",
    number = "26",
    pages = "261802",
    year = "2021"
}

@article{DarkSide-50:2022qzh,
    author = "Agnes, P. and others",
    collaboration = "DarkSide-50",
    title = "{Search for low-mass dark matter WIMPs with 12~ton-day exposure of DarkSide-50}",
    eprint = "2207.11966",
    archivePrefix = "arXiv",
    primaryClass = "hep-ex",
    reportNumber = "FERMILAB-PUB-22-589-ND-PPD-SCD",
    doi = "10.1103/PhysRevD.107.063001",
    journal = "Phys. Rev. D",
    volume = "107",
    number = "6",
    pages = "063001",
    year = "2023"
}

@article{LZ:2024zvo,
    author = "Aalbers, J. and others",
    collaboration = "LZ",
    title = "{Dark Matter Search Results from 4.2{\,}{\,}Tonne-Years of Exposure of the LUX-ZEPLIN (LZ) Experiment}",
    eprint = "2410.17036",
    archivePrefix = "arXiv",
    primaryClass = "hep-ex",
    reportNumber = "FERMILAB-PUB-24-0796-V",
    doi = "10.1103/4dyc-z8zf",
    journal = "Phys. Rev. Lett.",
    volume = "135",
    number = "1",
    pages = "011802",
    year = "2025"
}

@article{XENON:2025vwd,
    author = "Aprile, E. and others",
    collaboration = "XENON",
    title = "{WIMP Dark Matter Search using a 3.1 tonne $\times$ year Exposure of the XENONnT Experiment}",
    eprint = "2502.18005",
    archivePrefix = "arXiv",
    primaryClass = "hep-ex",
    month = "2",
    year = "2025"
}

@article{SuperCDMS:2020aus,
    author = "Alkhatib, I. and others",
    collaboration = "SuperCDMS",
    title = "{Light Dark Matter Search with a High-Resolution Athermal Phonon Detector Operated Above Ground}",
    eprint = "2007.14289",
    archivePrefix = "arXiv",
    primaryClass = "hep-ex",
    doi = "10.1103/PhysRevLett.127.061801",
    journal = "Phys. Rev. Lett.",
    volume = "127",
    pages = "061801",
    year = "2021"
}

@article{CRESST:2024cpr,
    author = "Angloher, G. and others",
    collaboration = "CRESST",
    title = "{First observation of single photons in a CRESST detector and new dark matter exclusion limits}",
    eprint = "2405.06527",
    archivePrefix = "arXiv",
    primaryClass = "astro-ph.CO",
    doi = "10.1103/PhysRevD.110.083038",
    journal = "Phys. Rev. D",
    volume = "110",
    number = "8",
    pages = "083038",
    year = "2024"
}

@article{SENSEI:2023zdf,
    author = "Adari, Prakruth and others",
    collaboration = "SENSEI",
    title = "{First Direct-Detection Results on Sub-GeV Dark Matter Using the SENSEI Detector at SNOLAB}",
    eprint = "2312.13342",
    archivePrefix = "arXiv",
    primaryClass = "astro-ph.CO",
    reportNumber = "YITP-SB-2023-30, FERMILAB-PUB-23-0824-CSAID-PPD",
    doi = "10.1103/PhysRevLett.134.011804",
    journal = "Phys. Rev. Lett.",
    volume = "134",
    number = "1",
    pages = "011804",
    year = "2025"
}

@article{Kouvaris:2016afs,
    author = "Kouvaris, Chris and Pradler, Josef",
    title = "{Probing sub-GeV Dark Matter with conventional detectors}",
    eprint = "1607.01789",
    archivePrefix = "arXiv",
    primaryClass = "hep-ph",
    doi = "10.1103/PhysRevLett.118.031803",
    journal = "Phys. Rev. Lett.",
    volume = "118",
    number = "3",
    pages = "031803",
    year = "2017"
}

@article{Bell:2023uvf,
    author = "Bell, Nicole F. and Cox, Peter and Dolan, Matthew J. and Newstead, Jayden L. and Ritter, Alexander C.",
    title = "{Exploring light dark matter with the Migdal effect in hydrogen-doped liquid xenon}",
    eprint = "2305.04690",
    archivePrefix = "arXiv",
    primaryClass = "hep-ph",
    doi = "10.1103/PhysRevD.109.L091902",
    journal = "Phys. Rev. D",
    volume = "109",
    number = "9",
    pages = "L091902",
    year = "2024"
}

@article{XENON:2019zpr,
    author = "Aprile, E. and others",
    collaboration = "XENON",
    title = "{Search for Light Dark Matter Interactions Enhanced by the Migdal Effect or Bremsstrahlung in XENON1T}",
    eprint = "1907.12771",
    archivePrefix = "arXiv",
    primaryClass = "hep-ex",
    doi = "10.1103/PhysRevLett.123.241803",
    journal = "Phys. Rev. Lett.",
    volume = "123",
    number = "24",
    pages = "241803",
    year = "2019"
}

@article{DarkSide:2022dhx,
    author = "Agnes, P. and others",
    collaboration = "DarkSide",
    title = "{Search for Dark-Matter{\textendash}Nucleon Interactions via Migdal Effect with DarkSide-50}",
    eprint = "2207.11967",
    archivePrefix = "arXiv",
    primaryClass = "hep-ex",
    doi = "10.1103/PhysRevLett.130.101001",
    journal = "Phys. Rev. Lett.",
    volume = "130",
    number = "10",
    pages = "101001",
    year = "2023"
}

@article{delfino2024multi,
  title={Multi-methodological analysis of hydrogen desorption from graphene},
  author={Delfino, Francesco and Ros, Carles and Palardonio, Sidney M. and Carretero, Nina M. and Murcia-L{\'o}pez, Sebasti{\'a}n and Morante, Juan Ram{\'o}n and Martorell, Jordi and Fthenakis, Zacharias G. and Sgroi, Mauro Francesco and Tozzini, Valentina and others},
  journal={Carbon},
  volume={227},
  pages={119211},
  year={2024},
  publisher={Elsevier},
  doi={10.1016/j.carbon.2024.119211}
}

@article{PTOLEMY:2022ldz,
    author = "Apponi, A. and others",
    collaboration = "PTOLEMY",
    title = "{Heisenberg{\textquoteright}s uncertainty principle in the PTOLEMY project: A theory update}",
    eprint = "2203.11228",
    archivePrefix = "arXiv",
    primaryClass = "hep-ph",
    doi = "10.1103/PhysRevD.106.053002",
    journal = "Phys. Rev. D",
    volume = "106",
    number = "5",
    pages = "053002",
    year = "2022"
}

@article{Casale:2025vgd,
    author = "Casale, Andrea and Esposito, Angelo and Menichetti, Guido and Tozzini, Valentina",
    title = "{{\ensuremath{\beta}}-decay spectrum of tritiated graphene: Combining nuclear quantum mechanics with density functional theory}",
    eprint = "2504.13259",
    archivePrefix = "arXiv",
    primaryClass = "hep-ph",
    doi = "10.1103/gr8x-lf9f",
    journal = "Phys. Rev. C",
    volume = "113",
    number = "5",
    pages = "054607",
    year = "2026"
}

@article{betti2022gap,
  title={Gap opening in double-sided highly hydrogenated free-standing graphene},
  author={Betti, Maria Grazia and Placidi, Ernesto and Izzo, Chiara and Blundo, Elena and Polimeni, Antonio and Sbroscia, Marco and Avila, Jos{\'e} and Dudin, Pavel and Hu, Kailong and Ito, Yoshikazu and others},
  journal={Nano letters},
  volume={22},
  number={7},
  pages={2971--2977},
  year={2022},
  publisher={ACS Publications},
  doi={10.1021/acs.nanolett.2c00162}
}

@book{giustino2014materials,
  title={Materials modelling using density functional theory: properties and predictions},
  author={Giustino, Feliciano},
  year={2014},
  publisher={Oxford University Press},
  url={https://global.oup.com/academic/product/materials-modelling-using-density-functional-theory-9780199662432?cc=it&lang=en&}
}

@article{kohn1965self,
  title={Self-consistent equations including exchange and correlation effects},
  author={Kohn, Walter and Sham, Lu Jeu},
  journal={Physical review},
  volume={140},
  number={4A},
  pages={A1133},
  year={1965},
  publisher={APS},
  doi={10.1103/PhysRev.140.A1133}
}

@article{PICO:2017tgi,
    author = "Amole, C. and others",
    collaboration = "PICO",
    title = "{Dark Matter Search Results from the PICO-60 C$_3$F$_8$ Bubble Chamber}",
    eprint = "1702.07666",
    archivePrefix = "arXiv",
    primaryClass = "astro-ph.CO",
    reportNumber = "FERMILAB-PUB-17-058-AE-PPD",
    doi = "10.1103/PhysRevLett.118.251301",
    journal = "Phys. Rev. Lett.",
    volume = "118",
    number = "25",
    pages = "251301",
    year = "2017"
}

@article{Apponi:2025fmu,
    author = "Apponi, Alice and others",
    title = "{Stability of Highly Hydrogenated Monolayer Graphene in Ultra-High Vacuum and in Air}",
    eprint = "2504.11853",
    archivePrefix = "arXiv",
    primaryClass = "cond-mat.mtrl-sci",
    month = "4",
    year = "2025"
}

@article{apponi2025highly,
  title={Highly Hydrogenated Monolayer Graphene with Wide Band Gap Opening},
  author={Apponi, Alice and Castellano, Orlando and Paoloni, Daniele and Convertino, Domenica and Mishra, Neeraj and Coletti, Camilla and Mariani, Carlo and Ruocco, Alessandro},
  eprint = {2504.10238},
  archivePrefix = "arXiv",
  primaryClass = "cond-mat.mtrl-sci",
  month = "4",
  year = "2025"
}

@article{abdelnabi2021deuterium,
  title={Deuterium adsorption on free-standing graphene},
  author={Abdelnabi, Mahmoud Mohamed Saad and Izzo, Chiara and Blundo, Elena and Betti, Maria Grazia and Sbroscia, Marco and Di Bella, Giulia and Cavoto, Gianluca and Polimeni, Antonio and Garc{\'\i}a-Cort{\'e}s, Isabel and Rucandio, Isabel and others},
  journal={Nanomaterials},
  volume={11},
  number={1},
  pages={130},
  year={2021},
  publisher={MDPI},
  doi={10.3390/nano11010130}
}

@article{tayyab2023spectromicroscopy,
  title={Spectromicroscopy study of induced defects in ion-bombarded highly aligned carbon nanotubes},
  author={Tayyab, Sammar and Apponi, Alice and Betti, Maria Grazia and Blundo, Elena and Cavoto, Gianluca and Frisenda, Riccardo and Jim{\'e}nez-Ar{\'e}valo, Nuria and Mariani, Carlo and Pandolfi, Francesco and Polimeni, Antonio and others},
  journal={Nanomaterials},
  volume={14},
  number={1},
  pages={77},
  year={2023},
  publisher={MDPI},
  doi={10.3390/nano14010077}
}

@article{Piffl:2013mla,
    author = "Piffl, Til and others",
    title = "{The RAVE survey: the Galactic escape speed and the mass of the Milky Way}",
    eprint = "1309.4293",
    archivePrefix = "arXiv",
    primaryClass = "astro-ph.GA",
    doi = "10.1051/0004-6361/201322531",
    journal = "Astron. Astrophys.",
    volume = "562",
    pages = "A91",
    year = "2014"
}

@article{monari2018escape,
    author = "Monari, G. and Famaey, B. and Carrillo, I. and Piffl, T. and Steinmetz, M. and Wyse, R. F. G. and Anders, F. and Chiappini, C. and Jan{\ss}en, K.",
    title = "{The escape speed curve of the Galaxy obtained from Gaia DR2 implies a heavy Milky Way}",
    eprint = "1807.04565",
    archivePrefix = "arXiv",
    primaryClass = "astro-ph.GA",
    doi = "10.1051/0004-6361/201833748",
    journal = "Astron. Astrophys.",
    volume = "616",
    pages = "L9",
    year = "2018",
    doi={10.1051/0004-6361/201833748}
}

@article{ou2024dark,
  title={The dark matter profile of the Milky Way inferred from its circular velocity curve},
  author={Ou, Xiaowei and Eilers, Anna-Christina and Necib, Lina and Frebel, Anna},
  journal={Monthly Notices of the Royal Astronomical Society},
  volume={528},
  number={1},
  pages={693--710},
  year={2024},
  publisher={Oxford University Press},
  doi={10.1093/mnras/stae034}
}

@article{SUNDARARAMAN2017278,
title = {JDFTx: Software for joint density-functional theory},
journal = {SoftwareX},
volume = {6},
pages = {278-284},
year = {2017},
issn = {2352-7110},
doi = {https://doi.org/10.1016/j.softx.2017.10.006},
url = {https://www.sciencedirect.com/science/article/pii/S2352711017300559},
author = {Ravishankar Sundararaman and Kendra Letchworth-Weaver and Kathleen A. Schwarz and Deniz Gunceler and Yalcin Ozhabes and Tomás A. Arias}
}

@article{PhysRevLett.77.3865,
  title = {Generalized Gradient Approximation Made Simple},
  author = {Perdew, John P. and Burke, Kieron and Ernzerhof, Matthias},
  journal = {Phys. Rev. Lett.},
  volume = {77},
  issue = {18},
  pages = {3865--3868},
  numpages = {0},
  year = {1996},
  month = {Oct},
  publisher = {American Physical Society},
  doi = {10.1103/PhysRevLett.77.3865},
  url = {https://link.aps.org/doi/10.1103/PhysRevLett.77.3865}
}

@article{SCHLIPF201536,
title = {Optimization algorithm for the generation of ONCV pseudopotentials},
journal = {Computer Physics Communications},
volume = {196},
pages = {36-44},
year = {2015},
issn = {0010-4655},
doi = {https://doi.org/10.1016/j.cpc.2015.05.011},
url = {https://www.sciencedirect.com/science/article/pii/S0010465515001897},
author = {Martin Schlipf and François Gygi}
}

@article{PhysRevB.87.165122,
  title = {Regularization of the Coulomb singularity in exact exchange by Wigner-Seitz truncated interactions: Towards chemical accuracy in nontrivial systems},
  author = {Sundararaman, Ravishankar and Arias, T. A.},
  journal = {Phys. Rev. B},
  volume = {87},
  issue = {16},
  pages = {165122},
  numpages = {13},
  year = {2013},
  month = {Apr},
  publisher = {American Physical Society},
  doi = {10.1103/PhysRevB.87.165122},
  url = {https://link.aps.org/doi/10.1103/PhysRevB.87.165122}
}

@article{Boehm:2003hm,
    author = "Boehm, C. and Fayet, Pierre",
    title = "{Scalar dark matter candidates}",
    eprint = "hep-ph/0305261",
    archivePrefix = "arXiv",
    doi = "10.1016/j.nuclphysb.2004.01.015",
    journal = "Nucl. Phys. B",
    volume = "683",
    pages = "219--263",
    year = "2004"
}

@article{Hooper:2008im,
    author = "Hooper, Dan and Zurek, Kathryn M.",
    title = "{A Natural Supersymmetric Model with MeV Dark Matter}",
    eprint = "0801.3686",
    archivePrefix = "arXiv",
    primaryClass = "hep-ph",
    reportNumber = "FERMILAB-PUB-07-587-A",
    doi = "10.1103/PhysRevD.77.087302",
    journal = "Phys. Rev. D",
    volume = "77",
    pages = "087302",
    year = "2008"
}

@article{Feng:2008ya,
    author = "Feng, Jonathan L. and Kumar, Jason",
    title = "{The WIMPless Miracle: Dark-Matter Particles without Weak-Scale Masses or Weak Interactions}",
    eprint = "0803.4196",
    archivePrefix = "arXiv",
    primaryClass = "hep-ph",
    reportNumber = "UCI-TR-2008-10",
    doi = "10.1103/PhysRevLett.101.231301",
    journal = "Phys. Rev. Lett.",
    volume = "101",
    pages = "231301",
    year = "2008"
}

@article{Falkowski:2011xh,
    author = "Falkowski, Adam and Ruderman, Joshua T. and Volansky, Tomer",
    title = "{Asymmetric Dark Matter from Leptogenesis}",
    eprint = "1101.4936",
    archivePrefix = "arXiv",
    primaryClass = "hep-ph",
    reportNumber = "LPT-ORSAY-11-09",
    doi = "10.1007/JHEP05(2011)106",
    journal = "JHEP",
    volume = "05",
    pages = "106",
    year = "2011"
}

@article{Hochberg:2014dra,
    author = "Hochberg, Yonit and Kuflik, Eric and Volansky, Tomer and Wacker, Jay G.",
    title = "{Mechanism for Thermal Relic Dark Matter of Strongly Interacting Massive Particles}",
    eprint = "1402.5143",
    archivePrefix = "arXiv",
    primaryClass = "hep-ph",
    doi = "10.1103/PhysRevLett.113.171301",
    journal = "Phys. Rev. Lett.",
    volume = "113",
    pages = "171301",
    year = "2014"
}

@article{DAgnolo:2015ujb,
    author = "D'Agnolo, Raffaele Tito and Ruderman, Joshua T.",
    title = "{Light Dark Matter from Forbidden Channels}",
    eprint = "1505.07107",
    archivePrefix = "arXiv",
    primaryClass = "hep-ph",
    doi = "10.1103/PhysRevLett.115.061301",
    journal = "Phys. Rev. Lett.",
    volume = "115",
    number = "6",
    pages = "061301",
    year = "2015"
}

@article{Kuflik:2015isi,
    author = "Kuflik, Eric and Perelstein, Maxim and Lorier, Nicolas Rey-Le and Tsai, Yu-Dai",
    title = "{Elastically Decoupling Dark Matter}",
    eprint = "1512.04545",
    archivePrefix = "arXiv",
    primaryClass = "hep-ph",
    doi = "10.1103/PhysRevLett.116.221302",
    journal = "Phys. Rev. Lett.",
    volume = "116",
    number = "22",
    pages = "221302",
    year = "2016"
}

@article{DAgnolo:2018wcn,
    author = "D'Agnolo, Raffaele Tito and Mondino, Cristina and Ruderman, Joshua T. and Wang, Po-Jen",
    title = "{Exponentially Light Dark Matter from Coannihilation}",
    eprint = "1803.02901",
    archivePrefix = "arXiv",
    primaryClass = "hep-ph",
    reportNumber = "CERN-TH-2018-032",
    doi = "10.1007/JHEP08(2018)079",
    journal = "JHEP",
    volume = "08",
    pages = "079",
    year = "2018"
}

@article{Essig:2011nj,
    author = "Essig, Rouven and Mardon, Jeremy and Volansky, Tomer",
    title = "{Direct Detection of Sub-GeV Dark Matter}",
    eprint = "1108.5383",
    archivePrefix = "arXiv",
    primaryClass = "hep-ph",
    reportNumber = "SLAC-PUB-14538",
    doi = "10.1103/PhysRevD.85.076007",
    journal = "Phys. Rev. D",
    volume = "85",
    pages = "076007",
    year = "2012"
}

@article{giannozz_01,
    author ={S. Baroni and S. de Gironcoli and A. Dal Corso and P. Giannozzi},
    collaboration = {},
    title = {Phonons and related crystal properties from density-functional
perturbation theory},
    eprint = {},
    archivePrefix = {},
    reportNumber = {},
    journal = {Rev Mod Phys},
    volume = "73",
    pages = "515--562",
    year = "2001",
    doi={10.1103/RevModPhys.73.515}
}

@article{Capparelli:2014lua,
    author = "Capparelli, L. M. and Cavoto, G. and Mazzilli, D. and Polosa, A. D.",
    title = "{Directional Dark Matter Searches with Carbon Nanotubes}",
    eprint = "1412.8213",
    archivePrefix = "arXiv",
    primaryClass = "physics.ins-det",
    doi = "10.1016/j.dark.2015.08.002",
    journal = "Phys. Dark Univ.",
    volume = "9-10",
    pages = "24--30",
    year = "2015",
    note = "[Erratum: Phys.Dark Univ. 11, 79--80 (2016)]"
}

@article{Cavoto:2016lqo,
    author = "Cavoto, G. and Cirillo, E. N. M. and Cocina, F. and Ferretti, J. and Polosa, A. D.",
    title = "{WIMP detection and slow ion dynamics in carbon nanotube arrays}",
    eprint = "1602.03216",
    archivePrefix = "arXiv",
    primaryClass = "physics.ins-det",
    doi = "10.1140/epjc/s10052-016-4193-7",
    journal = "Eur. Phys. J. C",
    volume = "76",
    number = "6",
    pages = "349",
    year = "2016"
}

@article{Cavoto:2017otc,
    author = "Cavoto, G. and Luchetta, F. and Polosa, A. D.",
    title = "{Sub-GeV Dark Matter Detection with Electron Recoils in Carbon Nanotubes}",
    eprint = "1706.02487",
    archivePrefix = "arXiv",
    primaryClass = "hep-ph",
    doi = "10.1016/j.physletb.2017.11.064",
    journal = "Phys. Lett. B",
    volume = "776",
    pages = "338--344",
    year = "2018"
}

@article{Hochberg:2016ntt,
    author = "Hochberg, Yonit and Kahn, Yonatan and Lisanti, Mariangela and Tully, Christopher G. and Zurek, Kathryn M.",
    title = "{Directional detection of dark matter with two-dimensional targets}",
    eprint = "1606.08849",
    archivePrefix = "arXiv",
    primaryClass = "hep-ph",
    doi = "10.1016/j.physletb.2017.06.051",
    journal = "Phys. Lett. B",
    volume = "772",
    pages = "239--246",
    year = "2017"
}

@article{Catena:2023awl,
    author = "Catena, Riccardo and Emken, Timon and Matas, Marek and Spaldin, Nicola A. and Urdshals, Einar",
    title = "{Direct searches for general dark matter-electron interactions with graphene detectors: Part II. Sensitivity studies}",
    eprint = "2303.15509",
    archivePrefix = "arXiv",
    primaryClass = "hep-ph",
    doi = "10.1103/PhysRevResearch.5.043258",
    journal = "Phys. Rev. Res.",
    volume = "5",
    number = "4",
    pages = "043258",
    year = "2023"
}

@article{Catena:2023qkj,
    author = "Catena, Riccardo and Emken, Timon and Matas, Marek and Spaldin, Nicola A. and Urdshals, Einar",
    title = "{Direct searches for general dark matter-electron interactions with graphene detectors: Part I. Electronic structure calculations}",
    eprint = "2303.15497",
    archivePrefix = "arXiv",
    primaryClass = "hep-ph",
    doi = "10.1103/PhysRevResearch.5.043257",
    journal = "Phys. Rev. Res.",
    volume = "5",
    number = "4",
    pages = "043257",
    year = "2023"
}

@article{PTOLEMY:2019hkd,
    author = "Betti, M. G. and others",
    collaboration = "PTOLEMY",
    title = "{Neutrino physics with the PTOLEMY project: active neutrino properties and the light sterile case}",
    eprint = "1902.05508",
    archivePrefix = "arXiv",
    primaryClass = "astro-ph.CO",
    doi = "10.1088/1475-7516/2019/07/047",
    journal = "JCAP",
    volume = "07",
    pages = "047",
    year = "2019"
}

@article{wahab2023proton,
  title={Proton transport through nanoscale corrugations in two-dimensional crystals},
  author={Wahab, Oluwasegun J and Daviddi, E. and Xin, B. and Sun, P. Z. and Griffin, E. and Colburn, A. W. and Barry, D. and Yagmurcukardes, M. and Peeters, F. M. and Geim, A. K. and others},
  journal={Nature},
  volume={620},
  number={7975},
  pages={782--786},
  year={2023},
  publisher={Nature Publishing Group UK London},
  doi={10.1038/s41586-023-06247-6}
}

@article{hu2014proton,
  title={Proton transport through one-atom-thick crystals},
  author={Hu, S. and Lozada-Hidalgo, M. and Wang, F. C. and Mishchenko, Artem and Schedin, F. and Nair, Rahul Raveendran andgHill, E. W. and Boukhvalov, D. W. and Katsnelson, M. I. and Dryfe, Robert A. W. and others},
  journal={Nature},
  volume={516},
  number={7530},
  pages={227--230},
  year={2014},
  publisher={Nature Publishing Group UK London},
  doi={10.1038/nature14015}
}

@article{griffin2020proton,
  title={Proton and Li-ion permeation through graphene with eight-atom-ring defects},
  author={Griffin, Eoin and Mogg, Lucas and Hao, Guang-Ping and Kalon, Gopinadhan and Bacaksiz, Cihan and Lopez-Polin, Guillermo and Zhou, T. Y. and Guarochico, Victor and Cai, Junhao and Neumann, Christof and others},
  journal={Acs Nano},
  volume={14},
  number={6},
  pages={7280--7286},
  year={2020},
  publisher={ACS Publications},
  doi={10.1021/acsnano.0c02496}
}

@article{lozada2018giant,
  title={Giant photoeffect in proton transport through graphene membranes},
  author={Lozada-Hidalgo, Marcelo and Zhang, Sheng and Hu, Sheng and Kravets, Vasyl G. and Rodriguez, Francisco J. and Berdyugin, Alexey and Grigorenko, Alexander and Geim, Andre K.},
  journal={Nature nanotechnology},
  volume={13},
  number={4},
  pages={300--303},
  year={2018},
  publisher={Nature Publishing Group UK London},
  doi={10.1038/s41565-017-0051-5}
}

@article{zeng2021biomimetic,
  title={Biomimetic N-doped graphene membrane for proton exchange membranes},
  author={Zeng, Zhiyang and Song, Ruiyang and Zhang, Shengping and Han, Xiao and Zhu, Zhen and Chen, Xiaobo and Wang, Luda},
  journal={Nano Letters},
  volume={21},
  number={10},
  pages={4314--4319},
  year={2021},
  publisher={ACS Publications},
  doi={10.1021/acs.nanolett.1c00813}
}

@article{tong2024control,
  title={Control of proton transport and hydrogenation in double-gated graphene},
  author={Tong, Jincheng and Fu, Yangming and Domaretskiy, Daniil and Della Pia, F. and Dagar, Parveen and Powell, Lewis and Bahamon, D. and Huang, Shiqi and Xin, Benhao and Costa Filho, R. N. and others},
  journal={Nature},
  volume={630},
  number={8017},
  pages={619--624},
  year={2024},
  publisher={Nature Publishing Group UK London},
  doi={10.1038/s41586-024-07435-8}
}

@article{horiguchi1978auger,
  title={Auger Neutralization of Slow Protons at Solid Surfaces},
  author={Horiguchi, S. and Koyama, K. and Ohtsuki, Y. H.},
  journal={physica status solidi (b)},
  volume={87},
  number={2},
  pages={757--763},
  year={1978},
  publisher={Wiley Online Library},
  doi={10.1002/pssb.2220870242}
}

@article{PandaX-4T_2023,
  title = {Search for Dark-Matter--Nucleon Interactions with a Dark Mediator in PandaX-4T},
  author = {Huang, Di and others},
  collaboration = {PandaX Collaboration},
  journal = {Phys. Rev. Lett.},
  volume = {131},
  issue = {19},
  pages = {191002},
  numpages = {6},
  year = {2023},
  month = {Nov},
  publisher = {American Physical Society},
  doi = {10.1103/PhysRevLett.131.191002},
  url = {https://link.aps.org/doi/10.1103/PhysRevLett.131.191002}
}

@article{Angloher:2025fzw,
    author = "Angloher, G. and others",
    title = "{The CRESST experiment: towards the next-generation of sub-GeV direct dark matter detection}",
    eprint = "2505.01183",
    archivePrefix = "arXiv",
    primaryClass = "astro-ph.CO",
    month = "5",
    year = "2025"
}

@inproceedings{SuperCDMS:2022kse,
    author = "Albakry, M. F. and others",
    collaboration = "SuperCDMS",
    title = "{A Strategy for Low-Mass Dark Matter Searches with Cryogenic Detectors in the SuperCDMS SNOLAB Facility}",
    booktitle = "{Snowmass 2021}",
    eprint = "2203.08463",
    archivePrefix = "arXiv",
    primaryClass = "physics.ins-det",
    reportNumber = "FERMILAB-CONF-22-171-PPD-SQMS",
    month = "3",
    year = "2022"
}

@article{TESSERACT:2025tfw,
    author = "Bui, T. K. and others",
    collaboration = "TESSERACT",
    title = "{First Limits on Light Dark Matter Interactions in a Low Threshold Two-Channel Athermal Phonon Detector from the TESSERACT Collaboration}",
    eprint = "2503.03683",
    archivePrefix = "arXiv",
    primaryClass = "hep-ex",
    doi = "10.1103/hsrl-crvf",
    journal = "Phys. Rev. Lett.",
    volume = "135",
    number = "16",
    pages = "161002",
    year = "2025"
}

@article{vonKrosigk:2022vnf,
    author = "von Krosigk, Belina and others",
    title = "{DELight: A Direct search Experiment for Light dark matter with superfluid helium}",
    eprint = "2209.10950",
    archivePrefix = "arXiv",
    primaryClass = "hep-ex",
    doi = "10.21468/SciPostPhysProc.12.016",
    journal = "SciPost Phys. Proc.",
    volume = "12",
    pages = "016",
    year = "2023"
}

@article{Colantoni:2020cet,
    author = "Colantoni, I. and others",
    title = "{BULLKID: BULky and Low-Threshold Kinetic Inductance Detectors}",
    doi = "10.1007/s10909-020-02408-3",
    journal = "J. Low Temp. Phys.",
    volume = "199",
    number = "3-4",
    pages = "593--597",
    year = "2020"
}

@article{Bento:2025ijg,
    author = "Bento, A. and others",
    title = "{The SWEET project: probing sugar crystals for direct dark matter searches}",
    eprint = "2510.00068",
    archivePrefix = "arXiv",
    primaryClass = "physics.ins-det",
    month = "9",
    year = "2025"
}

@article{Griffin:2024jec,
    author = "Griffin, Sin{\'e}ad M. and Hadas, Guy Daniel and Hochberg, Yonit and Inzani, Katherine and Lehmann, Benjamin V.",
    title = "{Dark-Matter{\textendash}Electron Detectors for Dark-Matter{\textendash}Nucleon Interactions}",
    eprint = "2412.16283",
    archivePrefix = "arXiv",
    primaryClass = "hep-ph",
    reportNumber = "MIT-CTP/5678",
    doi = "10.1103/6qqv-rl7q",
    journal = "Phys. Rev. Lett.",
    volume = "135",
    number = "14",
    pages = "141803",
    year = "2025"
}

@book{Migdal1977Qualitative,
  author       = {Migdal, Arkadii B.},
  title        = {Qualitative Methods in Quantum Theory},
  year         = {1977},
  publisher    = {W. A. Benjamin, Inc. (Advanced Book Program)},
  address      = {Reading, MA},
  series       = {Frontiers in Physics},
  volume       = {48},
  translator   = {Leggett, Anthony J.},
  isbn         = {0805370641},
  doi          = {10.1201/9780429497940}
}

@article{APPONI2026165658,
    author = "Apponi, Alice and others",
    title = "{Stability of highly hydrogenated monolayer graphene in ultra-high vacuum and in air}",
    eprint = "2504.11853",
    archivePrefix = "arXiv",
    primaryClass = "cond-mat.mtrl-sci",
    doi = "10.1016/j.apsusc.2025.165658",
    journal = "Appl. Surf. Sci.",
    volume = "723",
    pages = "165658",
    year = "2026",
    doi = {10.1016/j.apsusc.2025.165658}
}

@article{SIMSON2007772,
title = {Detection of low-energy protons using a silicon drift detector},
journal = {Nuclear Instruments and Methods in Physics Research Section A: Accelerators, Spectrometers, Detectors and Associated Equipment},
volume = {581},
number = {3},
pages = {772-775},
year = {2007},
issn = {0168-9002},
doi = {https://doi.org/10.1016/j.nima.2007.08.156},
url = {https://www.sciencedirect.com/science/article/pii/S0168900207018608},
author = {M. Simson and P. Holl and A.R. Müller and A. Niculae and G. Petzoldt and K. Schreckenbach and H. Soltau and L. Strüder and H.-F. Wirth and O. Zimmer}
}

@Article{D5NR02221E,
author ="Cecchini, Luca and Pepe, Carlo and Corcione, Benedetta and Castellano, Orlando and Paoloni, Daniele and Malnati, Federico and Cavoto, Gianluca and Carminati, Marco and Fiorini, Carlo and Pettinari, Giorgio and Yadav, Ravi Prakash and Rago, Ilaria and Apponi, Alice and Puiu, Andrei and Mariani, Carlo and Rajteri, Mauro and Ruocco, Alessandro and Pandolfi, Francesco",
title  ="Quantitative correlation between carbon nanotube tip morphology and field emission properties at cryogenic temperature",
journal  ="Nanoscale",
year  ="2025",
volume  ="17",
issue  ="36",
pages  ="21260-21267",
publisher  ="The Royal Society of Chemistry",
doi  ="10.1039/D5NR02221E",
url  ="http://dx.doi.org/10.1039/D5NR02221E"}

@book{Martin2004ElectronicStructure,
  author    = {Martin, Richard M.},
  title     = {Electronic Structure: Basic Theory and Practical Methods},
  publisher = {Cambridge University Press},
  year      = {2004},
  isbn      = {9780521782852},
  doi       = {10.1017/CBO9780511805769}
}

@article{Winter:Auger:1990,
  title = {Neutralization of fast protons in grazing collisions with a clean Al(111) surface},
  author = {Winter, H. and Kirsch, R. and Poizat, J. C. and Remillieux, J.},
  journal = {Phys. Rev. A},
  volume = {43},
  issue = {3},
  pages = {1660--1662},
  numpages = {0},
  year = {1991},
  month = {Feb},
  publisher = {American Physical Society},
  doi = {10.1103/PhysRevA.43.1660},
  url = {https://link.aps.org/doi/10.1103/PhysRevA.43.1660}
}

@article{Jouin:Auger:2011,
  title = {Velocity dependence of outgoing neutral fractions for H(1$s$) and H${}^{+}$ beams impinging on Al(111) at grazing incidence},
  author = {Jouin, H. and Gutierrez, F. A.},
  journal = {Phys. Rev. A},
  volume = {84},
  issue = {1},
  pages = {014901},
  numpages = {4},
  year = {2011},
  month = {Jul},
  publisher = {American Physical Society},
  doi = {10.1103/PhysRevA.84.014901},
  url = {https://link.aps.org/doi/10.1103/PhysRevA.84.014901}
}

@article{ZIMNY:Auger:1991,
title = {Interplay of resonant and Auger processes in proton neutralization after grazing surface scattering},
journal = {Surface Science},
volume = {255},
number = {1},
pages = {135-156},
year = {1991},
issn = {0039-6028},
doi = {https://doi.org/10.1016/0039-6028(91)90017-M},
url = {https://www.sciencedirect.com/science/article/pii/003960289190017M},
author = {R. Zimny and Z. L. Mišković and N. N. Nedeljković and Lj. D. Nedeljković}
}

@article{chen2014chemical,
  title={Chemical vapor deposition growth of single-walled carbon nanotubes with controlled structures for nanodevice applications},
  author={Chen, Yabin and Zhang, Jin},
  journal={Accounts of chemical research},
  volume={47},
  number={8},
  pages={2273--2281},
  year={2014},
  publisher={ACS Publications},
  doi={10.1021/ar400314b}
}

@article{kumar2010chemical,
  title={Chemical vapor deposition of carbon nanotubes: a review on growth mechanism and mass production},
  author={Kumar, Mukul and Ando, Yoshinori},
  journal={Journal of nanoscience and nanotechnology},
  volume={10},
  number={6},
  pages={3739--3758},
  year={2010},
  publisher={American Scientific Publishers},
  doi={10.1166/jnn.2010.2939}
}

@article{Pandolfi:2021tkx,
    author = "Pandolfi, F. and Apponi, A. and Cavoto, G. and Mariani, C. and Rago, I. and Ruocco, A.",
    title = "{The dark-PMT: a novel directional light Dark Matter detector based on vertically-aligned carbon nanotubes}",
    doi = "10.1088/1742-6596/2156/1/012051",
    journal = "J. Phys. Conf. Ser.",
    volume = "2156",
    pages = "012051",
    year = "2021"
}

@article{Chao:2021liw,
    author = "Chao, Wei and Jin, Mingjie and Peng, Ying-Quan",
    title = "{Directly detecting sub-MeV dark matter via 3-body inelastic scattering process}",
    eprint = "2109.14944",
    archivePrefix = "arXiv",
    primaryClass = "hep-ph",
    doi = "10.1103/PhysRevD.107.093009",
    journal = "Phys. Rev. D",
    volume = "107",
    number = "9",
    pages = "093009",
    year = "2023"
}

@article{Das:2023cbv,
    author = "Das, Anirban and Jang, Jiho and Min, Hongki",
    title = "{Sub-MeV dark matter detection with bilayer graphene}",
    eprint = "2312.00866",
    archivePrefix = "arXiv",
    primaryClass = "hep-ph",
    doi = "10.1103/PhysRevD.110.043020",
    journal = "Phys. Rev. D",
    volume = "110",
    number = "4",
    pages = "043020",
    year = "2024"
}

@article{NEWS-G:2024jms,
    author = "Arora, M. M. and others",
    collaboration = "NEWS-G",
    title = "{Search for Light Dark Matter with NEWS-G at the Laboratoire Souterrain de Modane Using a Methane Target}",
    eprint = "2407.12769",
    archivePrefix = "arXiv",
    primaryClass = "hep-ex",
    doi = "10.1103/PhysRevLett.134.141002",
    journal = "Phys. Rev. Lett.",
    volume = "134",
    number = "14",
    pages = "141002",
    year = "2025"
}

@article{HydroX:2025nxn,
    author = "Lippincott, W. H. and others",
    collaboration = "HydroX",
    title = "{HydroX, a light dark matter search with hydrogen-doped liquid xenon time projection chambers}",
    eprint = "2505.13402",
    archivePrefix = "arXiv",
    primaryClass = "hep-ex",
    reportNumber = "FERMILAB-PUB-25-0357-ETD",
    doi = "10.1038/s42005-025-02168-0",
    journal = "Commun. Phys.",
    volume = "8",
    number = "1",
    pages = "244",
    year = "2025"
}

@article{Amaro:2024vuk,
    author = "Amaro, F. D. and others",
    title = "{Secondary scintillation yield from GEM electron avalanches in He-CF$_4$ and He-CF$_4$-isobutane for CYGNO {\textemdash} Directional Dark Matter search with an optical TPC}",
    doi = "10.1016/j.physletb.2024.138759",
    journal = "Phys. Lett. B",
    volume = "855",
    pages = "138759",
    year = "2024"
}

@article{Cox:2024rew,
    author = "Cox, Peter and Dolan, Matthew J. and Wood, Joshua",
    title = "{New limits on light dark matter-nucleon scattering}",
    eprint = "2408.12144",
    archivePrefix = "arXiv",
    primaryClass = "hep-ph",
    doi = "10.1103/ww13-v14j",
    journal = "Phys. Rev. D",
    volume = "112",
    number = "11",
    pages = "115021",
    year = "2025"
}

@article{Cox:2025toz,
    author = "Cox, Peter and Dolan, Matthew J. and Ghosh, Avirup",
    title = "{Irreducible Constraints on Hadronically Interacting Sub-GeV Dark Matter}",
    eprint = "2512.20825",
    archivePrefix = "arXiv",
    primaryClass = "hep-ph",
    month = "12",
    year = "2025"
}

@article{Apponi:2022nxn,
    author = "Apponi, A. and Pandolfi, F. and Rago, I. and Cavoto, G. and Mariani, C. and Ruocco, A.",
    title = "{Absolute efficiency of a two-stage microchannel plate for electrons in the 30 - 900 eV energy range}",
    eprint = "2203.14658",
    archivePrefix = "arXiv",
    primaryClass = "physics.ins-det",
    doi = "10.1088/1361-6501/ac3d07",
    journal = "Measur. Sci. Tech.",
    volume = "33",
    pages = "025102",
    year = "2022"
}

@article{Wulandari:2003cr,
    author = "Wulandari, H. and Jochum, J. and Rau, W. and von Feilitzsch, F.",
    title = "{Neutron flux underground revisited}",
    eprint = "hep-ex/0312050",
    archivePrefix = "arXiv",
    doi = "10.1016/j.astropartphys.2004.07.005",
    journal = "Astropart. Phys.",
    volume = "22",
    pages = "313--322",
    year = "2004"
}

@article{Babenko:2007ss,
    author = "Babenko, V. A. and Petrov, N. M.",
    title = "{Determination of low-energy parameters of neutron-proton scattering on the basis of modern experimental data from partial-wave analyses}",
    eprint = "0704.1024",
    archivePrefix = "arXiv",
    primaryClass = "nucl-th",
    doi = "10.1134/S1063778807040072",
    journal = "Phys. Atom. Nucl.",
    volume = "70",
    pages = "669--675",
    year = "2007"
}

@article{Stewart:2006cr,
    author = "Stewart, D. Y and Harrison, P. F. and Morgan, B. and Ramachers, Yorck Alexander",
    editor = "Liss, T. M.",
    title = "{Radiation shielding for underground low-background experiments}",
    eprint = "nucl-ex/0607032",
    archivePrefix = "arXiv",
    doi = "10.1063/1.2402703",
    journal = "AIP Conf. Proc.",
    volume = "870",
    number = "1",
    pages = "568--571",
    year = "2006"
}

@article{ripanti2019polarization,
  title={Polarization effects of transversal and longitudinal optical phonons in bundles of multiwall carbon nanotubes},
  author={Ripanti, Francesca and D’Acunto, Giulio and Betti, Maria Grazia and Mariani, Carlo and Bittencourt, Carla and Postorino, Paolo},
  journal={The Journal of Physical Chemistry C},
  volume={123},
  number={32},
  pages={20013--20019},
  year={2019},
  publisher={ACS Publications},
  doi={10.1021/acs.jpcc.9b02638}
}

@article{Knapen:2017xzo,
    author = "Knapen, Simon and Lin, Tongyan and Zurek, Kathryn M.",
    title = "{Light Dark Matter: Models and Constraints}",
    eprint = "1709.07882",
    archivePrefix = "arXiv",
    primaryClass = "hep-ph",
    doi = "10.1103/PhysRevD.96.115021",
    journal = "Phys. Rev. D",
    volume = "96",
    number = "11",
    pages = "115021",
    year = "2017"
}

\end{document}